\documentstyle[aps,epsf,graphics,multicol]{revtex}
\newcommand{\xy}{\textsl{XY} }
\begin{document}
\draft
\preprint{SNUTP 01-xxx}
\title{Phase ordering on small-world networks with nearest-neighbor edges}
\author{H. Hong$^*$ and M.Y. Choi}
\address{Department of Physics, Seoul National University,
Seoul 151-747, Korea}
\author{Beom Jun Kim}
\address{Department of Molecular Science and Technology, Ajou 
University, Suwon 442-749, Korea}

\maketitle
\begin{abstract}
We investigate global phase coherence in a system of coupled oscillators on a
small-world networks constructed from a ring with nearest-neighbor edges. 
The effects of both thermal noise and quenched randomness on phase ordering 
are examined and compared with the global coherence in the corresponding \xy model 
without quenched randomness. 
It is found that in the appropriate regime phase ordering emerges at finite 
temperatures, even for a tiny fraction of shortcuts.
Nature of the phase transition is also discussed.
\end{abstract}


\begin{multicols}{2}
It is now known that diverse systems in nature may have the same topological 
structure as the {\em small-world networks}, first modeled by 
Watts and Strogatz (WS)~\cite{ref:WS}. 
A small-world network is usually constructed from a locally connected regular 
network with given interaction range $k$, where some of the edges are randomly 
``rewired'', creating long-range ``shortcuts''.
The small-world effect in such a model refers to
a high degree of clustering as well as the small characteristic path length 
$\ell\sim\log N / \log (2k{-}1)$, 
which is defined to be the average of the shortest distance between two vertices 
in the network of size $N$ (i.e., $N$ vertices)~\cite{ref:WS,ref:Watts,ref:Barrat}. 
So far, a major of studies have mostly paid attention to the geometrical properties 
described by the above two quantities. 
On the other hand, some recent studies have considered dynamical systems put on 
small-world networks~\cite{ref:Watts,ref:Barrat,ref:NHmodel,ref:XY,ref:HHong}, 
to find, e.g., faster propagation of information and better computational power.
Such studies of dynamical systems, which apparently have wide applicability 
in physics, computational science, and biological science,
have usually been performed on small-world networks with the interaction 
range $k\geq 2$. 
The original WS model is poorly defined for $k=1$ since the finite probability 
of isolated vertices induced by rewiring of the connections between the 
vertices yields divergence of the characteristic path length of the system. 
This difficulty may be circumvented by modifying the network construction method
in such a way that shortcuts are added without removing local edges~\cite{ref:Newman}. 
This construction method leads to an increase of the number of total edges in the 
system; this is in contrast with the WS model 
where the total number of edges is conserved. 

In this paper we study a set of nonlinear coupled oscillators on small-world 
networks constructed from rings with nearest neighbor edges ($k=1$), with 
attention to the emergence of long-range phase ordering as the number of shortcuts
is increased.  In particular, the effects of thermal noise together with 
quenched randomness on phase ordering are explored and the aspects of the 
global coherence are compared with those in the 
corresponding \xy model without quenched randomness.

The small-world network considered in this paper is constructed in the following way:
First, a one-dimensional regular network with nearest neighbor connections
between $N$ vertices is constructed, with the periodic boundary conditions imposed. 
Then shortcuts are added with the probability $P$, between 
randomly chosen pairs of vertices.  While local connections remain intact, 
more than one edge between any two vertices as well as any edge 
connecting a vertex to itself are prohibited, although such multi-connections 
and self-connections have been shown not to change the qualitative 
behavior of the WS model~\cite{ref:Newman}.
Throughout this paper the probability $P$ is defined to be the ratio of the 
number of total long-range connections (shortcuts) to the number of the 
total local edges in the network.

At each vertex of this small-world network is located an oscillator;
an edge connecting two vertices represents coupling between the
two oscillators at those two vertices.
Describing the state of the $i$th oscillator located on vertex $i$ by
its phase $\phi_i$, we write the set of equations of motion
governing the dynamics of the $N$ oscillator system $(i= 1,2,\cdots,N)$:
\begin{equation}\label{eq:model}
\dot{\phi_i }(t) = \omega_i - 
J \sum_{j \in \Lambda_i} \sin (\phi_i -\phi_j ) + \eta_i (t),
\end{equation}
where the intrinsic frequency $\omega_i$ of the $i$th oscillator are quenched 
random variables distributed according to, e.g., the Gaussian distribution 
with variance $\sigma^2$.  
The set $\Lambda_i$ denotes the neighborhood of 
vertex $i$, consisting of the vertices connected to vertex $i$ (via either 
local edges or shortcuts) with the coupling strength $J$.
The last term $\eta_i$ on the right-hand side represents thermal noise, 
i.e., independent white noise with zero mean and correlations
\[
\langle \eta_i(t) \eta_j(t') \rangle = 2T \delta_{ij}\delta(t-t'),
\]
where the noise level $T \,(\geq 0)$ may be regarded as the temperature of 
the system with the Boltzmann constant $k_B$ set equal to unity. 
When all the oscillators are identical with no quenched randomness ($\sigma^2 =0$), 
Eq.~(\ref{eq:model}) describes the system of classical \xy spins, for which 
the Hamiltonian is given by  
\begin{equation}
 H = - J \sum_{j \in \Lambda_i} \cos(\phi_i - \phi_j). 
\end{equation}
For example, $\phi_i$ may represent the angular direction of the two-dimensional 
spin or the phase of the superconducting order parameter at vertex $i$.
The investigation of the phase transition in the \xy model on a WS small-world 
network has revealed the presence of long-range order at finite temperatures, even
for a tiny fraction of shortcuts~\cite{ref:XY}. 
On the other hand, in the absence of the thermal noise $(T=0)$, 
Eq.~(\ref{eq:model}) describes the dynamics of a set of coupled oscillators, 
which has also been studied on a WS small-world network
and found to exhibit collective synchronization effectively~\cite{ref:HHong}. 
These two systems can be described by Eq.~(\ref{eq:model}) in the 
appropriate limits; 
in this respect Eq.~(\ref{eq:model}) provides a very interesting model, 
which allows to explore both the dynamic synchronization transition driven by 
quenched randomness and the thermodynamic phase transition by thermal noise together.
When the set $\Lambda_i$ includes the whole $N$ oscillators, Eq.~(\ref{eq:model}) 
describes the dynamics of the fully-connected oscillators with the normalized 
coupling strength $J/N$.  The phase synchronization in such a globally 
coupled system has been extensively studied~\cite{ref:synch}, 
and the increase of the critical coupling strength due to the thermal noise 
has been found~\cite{ref:Saka}. 

We here investigate the effects of quenched randomness together with thermal noise 
on the small-world network constructed as above. 
Collective behavior of the coupled oscillators on such a small-world network is 
conveniently described by the order parameter
\begin{equation} \label{eq:m}
 m \equiv \left[\left\langle \left|
                   \frac{1}{N}\sum_j e^{i\phi_j} \right| \right\rangle\right],
\end{equation}
where $\langle \cdots \rangle$ represents the thermal average with respect 
to the thermal noise.  For a small-world network, it is also necessary to take 
the average over different network realizations, denoted by $[\cdots]$. 
To compute the order parameter, 
we have integrated numerically the set of equations of motion given by 
Eq.~(\ref{eq:model}) through the use of Heun's method~\cite{ref:Heun} 
with the discrete time step $\Delta t =0.05$. 
Typically, while the equations have been integrated for
$N_t = 4\times 10^3$ time steps, the data from the first $N_t/2$ steps
have been discarded in measuring quantities of interest.
Both $\Delta t$ and $N_t$ have been varied to verify that the measured
quantities are precise enough and the networks of various sizes, up to $N=1600$,
have been considered.
For each network size, we have performed one hundred independent runs with
different configurations of the intrinsic frequencies as well as different 
network realizations, over which averages have been taken.

\begin{figure}
\vspace*{12.0cm}
\includegraphics{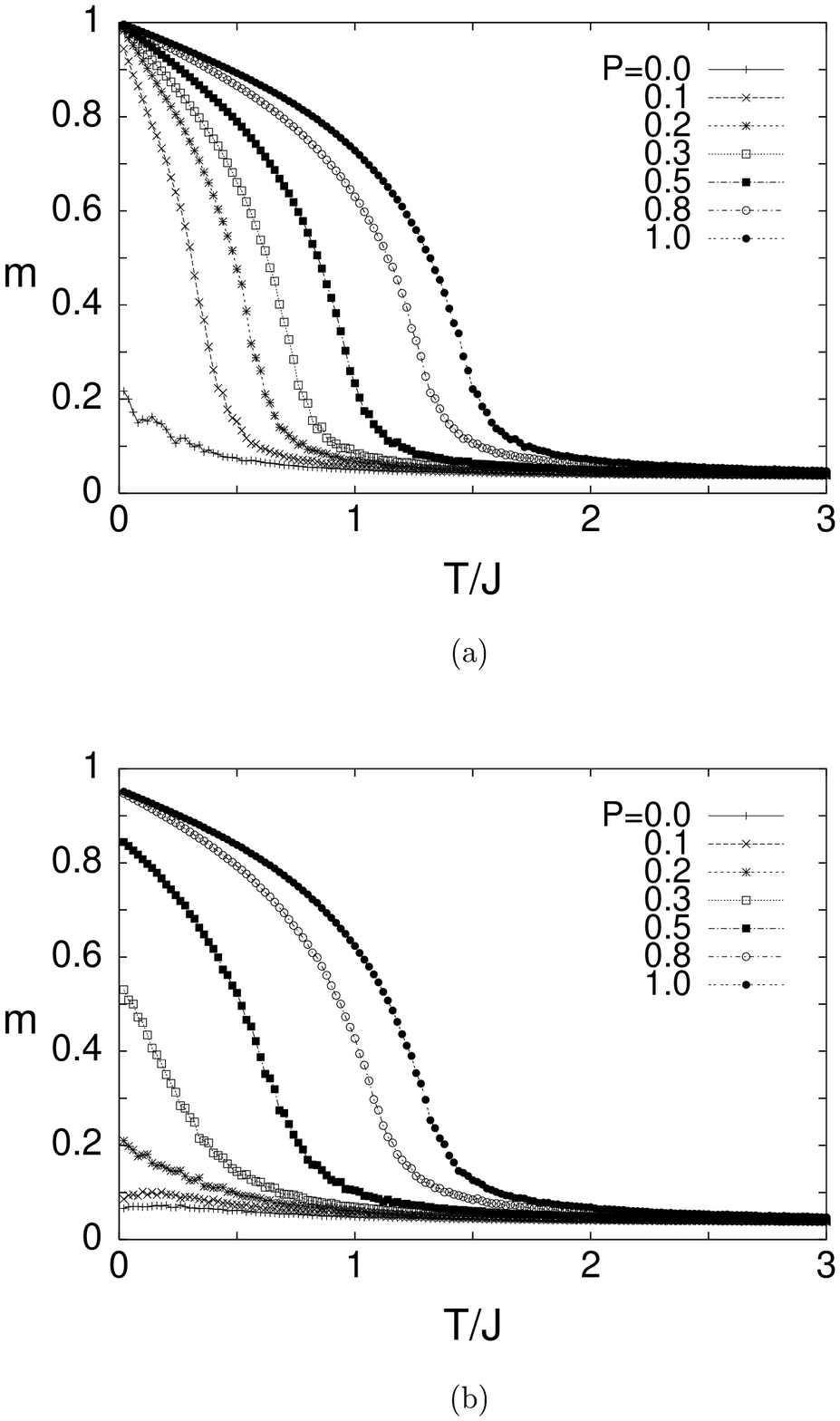}
\caption{Phase order parameter $m$ versus the rescaled temperature $T/J$ 
on the small-world network with the local interaction range $k=1$ 
and size $N=800$ for the variance (a) $\sigma^2 = 0.0$; (b) $\sigma^2 = 0.05$. 
The error bars estimated by the standard deviation have approximately the 
sizes of the symbols and the lines are merely guides to eyes. 
}
\label{fig:m_v0.05}
\end{figure}

Figure~\ref{fig:m_v0.05} displays the obtained order parameter $m$ 
versus the rescaled temperature $T/J$ at various values of the fraction $P$ 
on a network of size $N=800$
(a) in the absence of the quenched randomness ($\sigma^2=0$) 
and (b) in the presence of a finite amount of randomness ($\sigma^2=0.05$).
Thus Fig.~\ref{fig:m_v0.05}(a) exhibits the behavior of the magnetization 
with temperature in the corresponding \xy model. 
In the high-temperature limit $(T/J \rightarrow\infty$), phases of the oscillators 
are distributed uniformly on the interval $[0, 2\pi)$, leading to 
the absence of macroscopic coherence characterized by $m=O(N^{-1/2})$.  
In the opposite low-temperature limit $(T/J \rightarrow 0$), all 
the oscillators become synchronized for $P\neq 0$ to give $m=1$, 
regardless of the detailed structure of the network. 
On the other hand, Fig.~\ref{fig:m_v0.05}(b) indicates that the presence of 
randomness tends to suppress synchronization and macroscopic coherence 
is not achieved even at low temperatures for small but finite values of $P$. 
Indeed at low temperatures, the value of the order parameter for $P\lesssim 0.2$ in 
Fig.~\ref{fig:m_v0.05}(b) as well as that for $P=0$ in Fig.~\ref{fig:m_v0.05}(a) 
has been observed to diminish as the system size $N$ gets larger, 
suggesting the absence of coherence ($m=0$) in the thermodynamic limit. 
For $P\gtrsim 0.3$ in Fig.~\ref{fig:m_v0.05}(b), in contrast, the order parameter
at low temperatures tends to increase with the system size, indicating the 
emergence of synchronization. 

It is thus manifested in Fig.~\ref{fig:m_v0.05} that phase 
ordering exhibits strong dependence on the fraction $P$.  
Regardless of the quenched randomness, in particular, the phases
do not order in the absence of shortcuts ($P=0$), which is consistent 
with the known result in one dimension~\cite{ref:freqorder}. 
When some fraction of the shortcuts comes into the system, on the other hand, the 
dynamics of the system changes dramatically, giving rise phase ordering.
Here the critical value of the fraction $P_c$ beyond which the ordering 
emerges grows as the randomness is increased.  Namely, larger fractions 
of the long-ranged connections are needed for ordering in the system with 
stronger quenched randomness.  
For instance, Fig.~\ref{fig:m_v0.05}(a) shows the emergence of
ordering even for $P=0.1$ at low temperatures, which implies $P_c \lesssim 0.1$.
On the other hand, ordering in Fig.~\ref{fig:m_v0.05}(b) is observed 
for $P\gtrsim 0.3$ (but not for $P\lesssim 0.2$ even at low temperatures), 
apparently indicating that $0.2 \lesssim P_c \lesssim 0.3$.  
In this case of finite randomness ($\sigma^2=0.05$), 
extensive simulations yield $P_c \approx 0.23$ in the low-temperature limit. 
Note that the critical fraction $P_c$ in general increases with the variance 
$\sigma^2$ 
and the temperature $T$, reflecting that quenched randomness as well as 
thermal noise tends to suppress synchronization. 

To explore the nature of the phase transition and to determine accurately 
the critical temperature $T_c$, we examine the behavior of the order parameter 
by means of the finite-size scaling analysis. 
Recalling the critical behavior of the order parameter in the thermodynamic limit 
\begin{equation} \label{eq:mbeta}
m \sim (T_c - T )^{\beta} 
\end{equation}
with the critical exponent $\beta$, 
we expect the finite-size scaling form~\cite{ref:XY} 
\begin{equation} \label{eq:F}
m= N^{-\beta/{\bar\nu}} F\left((T{-}T_c)N^{1/{\bar\nu}}\right)
\end{equation}
with an appropriate scaling function $F$, where 
the critical exponent $\bar\nu$ describes the divergence of the
correlation volume $\xi_V$ at $T_c$~\cite{foot:xi,ref:xi}:
\begin{equation} \label{eq:correlation}
\xi_V \sim |T - T_c |^{-\bar\nu }.
\end{equation}
According to this scaling, the plot of $mN^{\beta/\bar\nu}$ versus $T$ 
should give a unique crossing point at $T_c$.  
After $\beta/{\bar \nu}$  and $T_c$ are determined from the plot 
of $mN^{\beta/{\bar\nu}}$ versus $T$, one then use
\begin{equation}
\ln\left[\frac{d m}{dT}\right]_{T_c} =  \frac{1-\beta}
{\bar\nu}\,\ln N + \mbox{const} ,
\label{eq:slope}
\end{equation}
to obtain the value of $(1-\beta)/{\bar \nu}$;
this, combined with the known value of $\beta/{\bar \nu}$, finally gives the values of
$\beta$ and ${\bar\nu}$. 

\begin{figure}
\resizebox*{!}{5.5cm}{\includegraphics{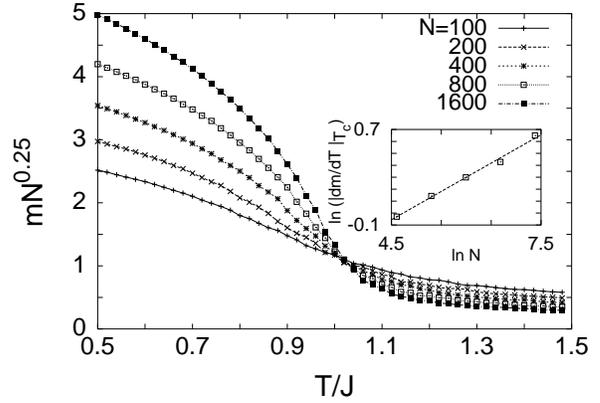}}
\caption{
Finite-size scaling of the phase order parameter [see Eq.~(\ref{eq:F})] 
on the small-world network with $k=1$ and $P=0.5$, 
in the absence of quenched randomness ($\sigma^2 =0$).
There exists a unique crossing point at $T_c \approx 1.02$.
Inset gives the slope $(1-\beta)/{\bar \nu} \approx 0.24$, which, 
combined with $\beta/\bar\nu \approx 0.25$ found in the main panel, results 
in $\beta \approx 0.51$ and $\bar\nu \approx 2.04$. 
} 
\label{fig:v0_P0.5}
\end{figure}

Figure~\ref{fig:v0_P0.5} shows the determination of $T_c$ on the 
small-world network with $k=1$ and $P=0.5$ in the absence of the 
quenched randomness ($\sigma^2 = 0$). 
Varying the value of $\beta/\bar\nu$, we find that $\beta/\bar\nu \approx 0.25$
gives the well-defined crossing point at $T_c \approx 1.02$ 
(in units of $J$). 
In the inset of Fig.~\ref{fig:v0_P0.5}, the least-square fit to Eq.~(\ref{eq:slope})
gives $(1-\beta)/\bar\nu \approx 0.24$, 
which, combined with $\beta/\bar\nu \approx 0.25$,
yields $\beta \approx 0.51$ and $\bar \nu \approx 2.04$.
These results demonstrate that the small-world network constructed from rings 
with nearest neighbor edges is also a mean-field system, which has the 
values $\beta = 1/2$ and $\nu =1/2$~\cite{ref:synch}.  
Further, the obtained value of $\bar\nu$ close to $2$ indicates that 
the upper critical dimension of the phase ordering transition is four~\cite{foot:xi}, 
the same as that on the usual small-world network of the WS type~\cite{ref:XY}. 

\begin{figure}
\resizebox*{!}{5.5cm}{\includegraphics{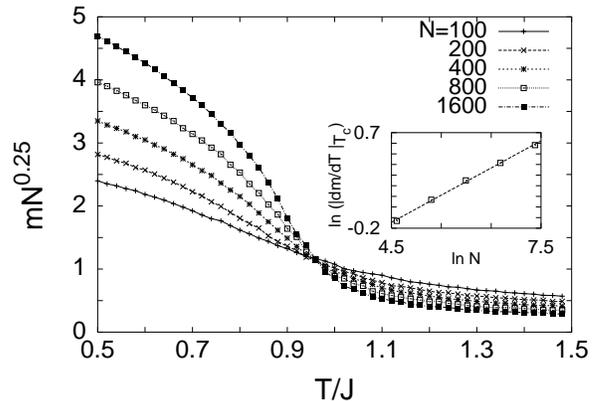}}
\caption{
Finite-size scaling of the phase order parameter
on the small-world network with $k=1$ and $P=0.5$ 
in the presence of a finite amount of randomness ($\sigma^2 =0.01$).
There is given a unique crossing point at $T_c \approx 0.96$, which is 
apparently smaller than $T_c \approx 1.02$ for $\sigma^2=0$.
Inset: The slope $(1-\beta)/{\bar \nu} \approx 0.26$ is obtained, which, 
combined with $\beta/\bar\nu \approx 0.25$, yields 
$\beta \approx 0.49$ and $\bar\nu \approx 1.96$.
} 
\label{fig:v0.01_P0.5}
\end{figure}

Similarly, the scaling form in the presence of finite randomness is 
plotted in Fig.~\ref{fig:v0.01_P0.5}, displaying the determination of $T_c$ and 
exponents for $\sigma^2 = 0.01$. 
Via the same analysis as in Fig.~\ref{fig:v0_P0.5}, we obtain 
$T_c \approx 0.96$ (again in units of $J$), which is smaller than the value in the 
system without quenched randomness. 
Thus the quenched randomness tends to suppress phase ordering, 
lowering the critical temperature. 
The corresponding values of the exponents $\beta \approx 0.49$ and 
$\bar\nu \approx 1.96$ are 
essentially the same as those in the absence of randomness ($\sigma^2=0$).
This concludes that the coupled oscillator system on a small-world network
with the number of connections given by ${\cal O}(N)$ displays 
a mean-field synchronization transition even for the shortest 
local-interaction range $k=1$, similarly to the system on a globally 
connected network with the much larger number of connections 
${\cal O}(N^2)$. 

\begin{figure}
\resizebox*{!}{11.0cm}{\includegraphics{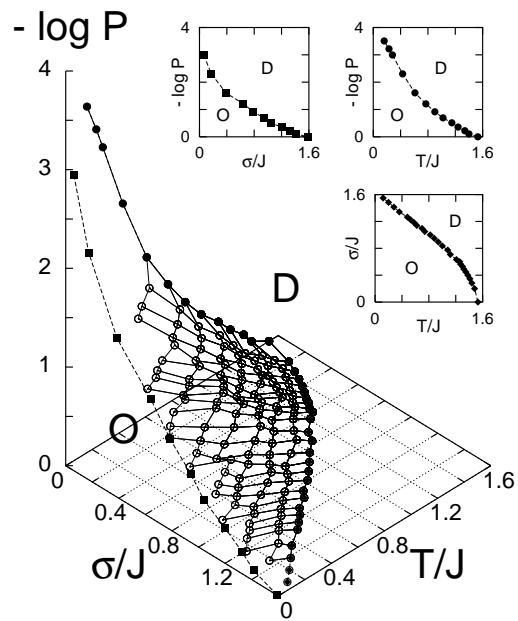}}
\caption{
Phase diagram of the oscillator system on a small-world network.  
The data points on the phase boundary have been obtained from the 
finite-size scaling of the order parameter in Eq.~(\ref{eq:F}).  
O and D represent the ordered state and the disordered one, respectively.
Insets: Phase boundaries on the planes $T=0$, $\sigma=0$, and 
$P=1$, respectively.}
\label{fig:phd}
\end{figure}

Figure~\ref{fig:phd} displays the phase diagram in the space of the 
rescaled standard deviation $\sigma/J$, the rescaled temperature 
$T/J$, and the logarithmic fraction $-\log P$.  
The data for the boundary surface separating ordered (O) and disordered (D) 
phase have been obtained from the finite-size scaling 
analysis, with the temperature $T$ varied for given variance 
$\sigma^2$ and the fraction $P$. 
In the absence of both quenched randomness and thermal noise ($\sigma^2=T=0$), 
the frequencies of the oscillators are already ordered even without any 
long-range connections, and the phases can be ordered through the 
coupling between the oscillators. 
Therefore we have the critical fraction $P_c =0$ in this case, 
which is implied in the first and the second insets of Fig.~\ref{fig:phd}. 
The third inset displays the boundary on the plane of $P=1$, 
which exhibits the interplay of the two types of the randomness 
(thermal noise and quenched randomness) in suppressing the phase ordering. 

In summary, we have examined phase ordering in the system of
coupled oscillators on small-world networks constructed from rings with 
nearest-neighbor edges.
The phase ordering has been found to depend strongly upon the 
fraction $P$ of the long-range connections to the short-range local ones. 
In particular, such ordering, which is absent at $P=0$ for nonzero 
quenched randomness ($\sigma^2 \neq 0$), emerges at finite 
temperatures as $P$ is increased from zero. 
The critical value $P_c$, below which ordering 
does not set in, apparently vanishes in the limits $T\rightarrow 0$ 
and $\sigma^2 \rightarrow 0$, increasing from zero with the randomness. 
The phase boundary between the ordered state and the disordered one, 
obtained from the finite-size scaling analysis, 
has been found to be of the mean-field type. 
Finally, we note that the disordered state in Fig.~\ref{fig:phd} represents 
phase disorder, thus the possibility of a different type of order 
such as frequency order may not be excluded, the investigation 
of which is left for further study. 

This work was supported in part by the Ministry of Education of Korea through the BK21 
Program (H.H. and M.Y.C.), and in part by the Swedish Natural Research Council through 
Contract No. F 5102-659/2001 (B.J.K.).

\end{multicols}

\begin{references}
\bibitem{foot:xi}
In $d$ dimensions, we have $\xi_{V} \sim \xi^d $ with the correlation length
diverging as $\xi \sim |K - K_c |^{-\nu}$.
This leads to the relation $\bar\nu = \nu d$ in a $d$-dimensional system;
in a mean-field system ($\nu =1/2$), we have $\bar\nu = d_c/2$ with the
upper critical dimension $d_c$.
\end{references}
\end{document}